\documentstyle[12pt]{article}
\def\ie{{\em i.e.}}
\def\eg{{\em e.g.}}

\def\beq{\begin{equation}}
\def\eeq{\end{equation}}

\catcode`\@=11 

\def\VEV#1{\left\langle #1\right\rangle}

\def\lsim{\mathrel{\mathpalette\@versim<}}
\def\gsim{\mathrel{\mathpalette\@versim>}}
\def\@versim#1#2{\vcenter{\offinterlineskip
    \ialign{$\m@th#1\hfil##\hfil$\crcr#2\crcr\sim\crcr } }}
\def\etal{{\em et. al.}}
\def\JL{J. L. Lopez}
\def\DVN{D. V. Nanopoulos}
\def\AZ{A. Zichichi}

\def\t1{{\tilde 1}}

\def\GeV{\,{\rm GeV}}
\def\TeV{\,{\rm TeV}}

\def\to{\rightarrow}

\def\NPB#1#2#3{Nucl. Phys. B {\bf#1} (19#2) #3}
\def\PLB#1#2#3{Phys. Lett. B {\bf#1} (19#2) #3}

\def\PRD#1#2#3{Phys. Rev. D {\bf#1} (19#2) #3}
\def\PRL#1#2#3{Phys. Rev. Lett. {\bf#1} (19#2) #3}

\begin{document}
\begin{flushright}
\baselineskip=12pt
{CERN-TH.7423/94}\\
{CTP-TAMU-59/94}\\
{ACT-17/94}\\
{hep-ph/9411281}
\end{flushright}

\begin{center}
{\Large\bf The march towards no-scale supergravity\\}
\vglue 0.5cm
{\large D.V. Nanopoulos}
\vglue 0.5cm
{Center for Theoretical Physics, Department of Physics, Texas A\&M
University\\}
{College Station, TX 77843--4242, USA\\}
{Astroparticle Physics Group, Houston Advanced Research Center
(HARC)\\}
{The Mitchell Campus, The Woodlands, TX 77381, USA\\}
{CERN Theory Division, 1211 Geneva 23, Switzerland}
\end{center}

\begin{abstract}
The different steps that led us to the discovery of {\em no-scale
supergravity} are discussed from a very personal point of view. No-scale
supergravity has been heralded as the most promising effective theory that
describes physics below the Planck scale. In its string-derived form it holds
the potential for a Dynamical Determination Of Everything (DDOE).
\end{abstract}

\baselineskip=14pt

\section{Overview}
Among the many problems facing particle physics today, two fundamental ones
pronouncedly stick out: (i) the origin of the different mass scales observed
in Nature, and (ii) the miniscule upper bound on the cosmological constant.
While, at first sight, these two problems seem to be unrelated, I will argue
below that there is a deep-rooted correlation that suitably exploited may lead
us to a natural and common resolution of these conundrums.
\footnotetext{Invited talk presented at the International Conference on:
``The History of Original Ideas and Basic Discoveries in Particle Physics",
Erice, Italy, 29 July - 4 August 1994.}

The success of the electroweak unification is based on the idea of Spontaneous
Symmetry Breaking (SSB) of certain gauge symmetries, by allowing some scalar
field ($\phi$) to get a vacuum expectation value (v.e.v.), such that
$\VEV{\phi}=v\approx G^{-1/2}_F$, with $G_F$ the Fermi constant. The SSB is
usually achieved by some judicious choice of the effective potential $V_{\rm
eff}(\phi)$, completely arbitrary at this level of sophistication. All of the
masses are then proportional to $v$, with the proportionality coefficient
depending on the nature of the particle under consideration: gauge bosons
$(W,Z)$: $m_{W,Z}\approx gv$, with $g$ some electroweak gauge coupling;
fermions: $m_{f_i}\approx h_i v$, with $h_i$ some arbitrary Yukawa coupling,
and for the Higgs boson: $m_{\phi}\approx \lambda^{1/2} v$, with $\lambda$
some arbitrary quartic coupling. All sounds well, as we seem to have identified
the origin of the different mass scales, \ie, spontaneous breakdown
($\VEV{\phi}\not=0$). This naive view turned out to be very deceptive. Each of
the above masses ($m_{W,Z},m_{f_i},m_\phi$) is the source of a cumbersome
problem. To start with, who ``orders" the Higgs potential to have the familiar
``dumbbell" shape, so that $\VEV{\phi}\not=0$? Furthermore, as it is
well-known, the quadratic ultraviolet divergences that plague scalar field
theories shift the Higgs mass at the one-loop level by an amount
\begin{equation}
\delta m^2_\phi\sim\Lambda^2\sim M^2_{Pl},
\label{1}
\end{equation}
where $M_{Pl}=(G_N)^{-1/2}$ is the natural cut-off of the theory, with $G_N$
Newton's gravitional constant, and assuming for simplicity that no intermediate
mass scales are present. Clearly, a catastrophic result, the {\em stability}
part of the {\em gauge hierarchy problem}. But even if we stabilized $m_\phi$,
why is it that
\begin{equation}
\left({G_N\over G_F}\right)^{1/2}\sim {M_W\over M_{Pl}}\sim10^{-17},
\label{2}
\end{equation}
\ie, the {\em magnitude} part of the {\em gauge hierarchy problem}. In
addition, a satisfactory resolution of the two-part gauge hierarchy problem
will, in principle, leave untouched the {\em fermion mass hierarchy problem},
since
\begin{equation}
h_i:\quad 10^{-6}\ (``e")\longrightarrow {\cal O}(1)\ (``t")
\label{3}
\end{equation}
a rather spread out fermion mass spectrum.

The SSB idea, while unavoidable for getting a {\em renormalizable} electroweak
theory, consistent with all presently available experimental data, not only
leaves us a bit unsatisfied about the ``origin of mass", but it is also mainly
responsible for the other fundamental problem, that of the cosmological
constant ($\Lambda_c$). A SSB vacuum would have a tremendously large energy
density and this would give the physical vacuum an enormously large
cosmological constant! It is fair to say that in todays popular theories, which
extend the SSB ideas far beyond the electroweak scale, it is difficult to
swallow the vast disparity between the ``expected" ${\cal O}(M^4_{Pl})$ and
the ``observed" upper bound on the cosmological constant
\begin{equation}
{\Lambda_c\over M^4_{Pl}}\le{\cal O}(10^{-120}).
\label{4}
\end{equation}
This is the notorious cosmological constant problem, sometimes called,
justifiably I think, the worst fine-tuning problem in the history of physics!

While SSB seems to be the common root of these two fundamental problems, we
should not forget that both are related in a big way with gravity, which until
now has been left out of our discussion. After all, another name for the mass
would be {\em gravitational charge}, and the SSB vacuum energy metamorphises
into the cosmological constant only when we couple our particle theory to
gravity. We should then expect that the search for a possible common solution
to these two problems has to involve gravity in a big way. This is exactly
what happens in {\bf no-scale supergravity}, which was discovered in 1983
\cite{1,2,3,4} in order to provide a common solution to the two fundamental
problems, by exploiting their common origin and suggesting a very deep
connection:
\begin{equation}
{\rm Existence\ of\ multitude\ of\ mass\ scales\longleftrightarrow Absence\ of\
cosmological\ constant}
\label{5}
\end{equation}
which may be considered as the ``signature" of no-scale supergravity
\cite{1}--\cite{6}.

In the next sections I will provide my personal account of the different steps
that brought us to the discovery of no-scale supergravity, and discuss its
impact and its present status. This talk is by no means a complete review
\cite{7} of no-scale supergravity, neither of supersymmetry or supergravity,
but some personal reminiscences of how things happened. Nevertheless, I do hope
to convince the reader of why no-scale supergravity has been heralded by many
\cite{8}, and not only by some of its inventors, as {\em a very good candidate}
for an effective theory explaining physics below the Planck scale!

\section{Applied supergravity}
Among the many spectacular properties of supersymmetry (SUSY) or fermion-boson
symmetry, the non-renormalization theorems play a fundamental role in solving
the stability part of the gauge hierarchy problem \cite{9}. Indeed, in an
exactly supersymmetric world, the Higgs mass would receive no radiative
corrections due to the fermion-boson loop cancellation. But exact supersymmetry
implies the equality between fermion and boson masses, not observed in Nature!
In more realistic, {\em broken} SUSY theories, one replaces (\ref{1}) by
\begin{equation}
\delta m^2_\phi=\Lambda^2=m^2_B-m^2_F\lsim{\cal O}(1\TeV^2)
\label{6}
\end{equation}
where the cut-off is now provided by the $(\rm mass)^2$ difference between the
relevant bosons and fermions. If and only if it happens that these $(\rm
mass)^2$ differences are smaller than ${\cal O}(1\TeV^2)$, we would get a
satisfactory resolution \cite{9} of the stability part of the gauge hierarchy
problem. Of course, in such case we effectively double the particle content of
our theory, so that for each particle, its superpartner is available, and with
a mass not too far from the Fermi scale ($G^{-1/2}_F$). The existence of an
experimentally accessible superworld \cite{10} is undoubtely the most dramatic
consequence of invoking SUSY to resolve the stability part of the gauge
hierarchy problem.

Thanks to the efforts of many people and many papers later \cite{7}, we
realized that the notion of a {\em realistic}, {\em spontaneously broken} SUSY
theory was in illusion. The notion of {\em softly broken} SUSY had to be
introduced \cite{11}, which despite its {\em arbitrariness} and {\em
ad-hoc-ness} seems to keep unscathed the elements of SUSY relevant to the
solution of the stability problem \cite{12}. Alas, the proliferation of the new
arbitrary parameters that had to be introduced made the predictability of the
theory mute. For every chiral multiplet, like say the electron (spin 1/2) --
selectron (spin 0) one throws in some mass parameter $m_0$, ditto for the gauge
multiplet, say the photon (spin 1) -- photino (spin 1/2) one introduces a mass
parameter $m_{1/2}$, while for the Yukawa couplings one has to introduce some
SUSY breaking parameter $A$ \cite{7,10}. Furthermore, one has to introduce at
least two Higgs doublets $(\phi_t,\phi_b)$, one coupled to the $Q=2/3$ quarks
and the other to the $Q=-1/3$ quarks  and $Q=-1$ leptons, thus introducing a
new angle $\beta$: $\tan\beta\equiv\VEV{\phi_t}/\VEV{\phi_b}=v_t/v_b$, and a
new mass parameter $\mu$: $\mu\phi_t\phi_b$, as well as its SUSY breaking
counterpart $B$ \cite{7,10}. Counting the whole lot we arrive at 27 new
parameters, not exactly a very desirable situation \cite{10}. Actually, Low
Energy Phenomenology (LEP) puts a lot of severe constraints on these
parameters, that may be represented {\em loosely} as follows:
\begin{itemize}
\item Absence, at lowest order, of FCNC: ``universal" $m_0$ \cite{13}.
\item Existence of gauge coupling unification at super high energies, as
``observed" at LEP \cite{14}: ``universal" $m_{1/2}$
\item Absence of a substantial neutron electron dipole moment $d_n\le{\cal
O}(10^{-25}{\rm e-cm})$: ``real" $A,B$ \cite{15}
\item and of course {\em stability}: $m_0,m_{1/2},\mu\lsim{\cal O}(1\TeV)$
\end{itemize}
\begin{equation}
\label{7}
\end{equation}

While all the above constraints can be met in the framework of softly broken
SUSY theories, its arbitrariness and lack of dynamical reasoning makes one
think that something better has to be invented. Indeed, until now we have
considered only rigid or global SUSY, by tacitly ignoring gravity. While in
the case of usual particle symmetries, the notion of global versus local is
a matter of taste, in the case of SUSY theories the mere existence of gravity
implies that SUSY theories are {\em local}. In other words, we have to address
from the start locally supersymmetric theories or {\em supergravity} (SUGRA)
theories \cite{7}. In the case of {\em locally} supersymmetric theories, like
in usual gauge theories, the only way to break them, at least in perturbation
theory, is {\em spontaneously}.

One of the basic steps towards the construction of a realistic SUGRA model
was taken by Arnowitt, Chamseddine, and Nath \cite{16} and independently by
Barbieri, Ferrara, and Savoy \cite{17}, when they proved that a spontaneously
broken SUGRA theory reduces at low energies (with respect to the Planck scale
$M_{Pl}$) to a global SUSY theory plus some ``soft breaking" terms! Exactly
what ``the doctor ordered"
\begin{equation}
{\cal L}_{\rm SUGRA}\ {\buildrel{\rm SSB\ at\ M_{Pl}}\over{---\longrightarrow}}
\ {\cal L}_{\rm SUSY}\oplus ``\rm soft\ breaking"\ terms\ ,
\label{8}
\end{equation}
where in this case all the soft breaking terms $(m_0,m_{1/2},A,B)$ are
functions of the ``hidden sector" fields responsible for the SSB of
supergravity
\begin{equation}
\widetilde m=\widetilde m(\VEV{T_i})\ .
\label{9}
\end{equation}
In (\ref{9}) I have generally represented the soft breaking parameters by
$\widetilde m$ and $\VEV{T_i}$ denote the hidden sector fields $T_i$ at their
corresponding minima. Thus, in this new framework the soft breaking parametes
may be viewed as boundary conditions, ``given" by the vacuum expectation values
of the corresponding functions at some energy scale close to $M_{Pl}$. The
gravitational origin of the soft breaking parameters, make the phenomenological
constraints (\ref{7}), demanding ``universal" values, easier to satisfy.
After all, as far as we know, gravity is blind to flavor, or $SU(2)\times U(1)$
or $SU(3)_{\rm color}$ quantum numbers, so it makes a lot of sense to assume a
common $m_0$ for all squarks and sleptons, and a common $m_{1/2}$ for all
gauginos, as a boundary condition close, or at, the Planck scale. Of course,
at lower energies these mass parameters, like all other mass parameters in
Quantum Field Theory, are subject to renormalizations, due to controlable
electroweak and strong interaction radiative corrections, as embodied in their
corresponding renormalization group equations (RGEs) \cite{7}.

Before going any further, it is worth emphasizing that the search for a
resolution of the stability part of the gauge hierarchy problem has led us to
a practical use of SUSY theories, which until then they were kind of
interesting field theories with some curious properties, like the
non-renormalization theorems! Furthermore, it was found that for practical
reasons, we had to include gravity, \ie, deal with supergravity theories,
which we ought to have used anyway. It is fair to say that the resolution
of the stability part of the gauge hierarchy problem has made SUSY/SUGRA
theories not only mathematically elegant, but most importantly, perhaps, also
physically relevant.

\section{Radiative Electroweak Symmetry Breaking (REWSB)}
One late afternoon, in the summer of `82, John Ellis, Kyriakos Tamvakis, and
myself were bouncing off ideas about how to use best the emerging framework
of applied supergravity. The atmosphere was kind of grim, since we had noticed
that the SUGRA-type ``universality" of the soft breaking terms for squarks,
sleptons, {\em and} Higgs bosons: $m^2_0\ge0$, was bad news for phenomenology.
The dilemma we were facing was the following: clearly $SU(3)_{\rm color}$ and
$U(1)_{\rm em}$ are unbroken symmetries and it is good that for squarks and
sleptons $m^2_0>0$, \ie, no potential danger for spontaneous breakdown, where
$m^2_0<0$ is needed. On the other hand, that is kind of bad for the
electroweak interactions, where SSB entails $m^2_0<0$! What is going on?
A grave problem. We decided to call it a day, and as I was walking back to my
office it occurred to me that we had been very naive. We had forgotten about
renormalization of these mass parameters! Supergravity just provides
$m^2_0\ge0$ at, or close to, the Planck scale, but then the different masses
are renormalized differently, depending on the strong and electromagnetic
``charges" of the corresponding particles.

It is basically the same type of idea that enabled Andrei Buras, John Ellis,
Mary K. Gaillard, and myself \cite{18}, a few years before (in `77) to
calculate successfully the $m_b/m_\tau$ ratio in GUTs and make a strong
prediction for the number of generations: $N_g=3$ \cite{18,19}. Twelve years
later this was successfully verified at LEP \cite{20}. Again there, one started
with a universal Yukawa coupling at the GUT scale $h_b=h_\tau$, and let mainly
strong and electroweak radiative corrections to provide the ``right" number at
low energies $m_b/m_\tau\approx3$.

I ran back to John's office and told them my thought, and I immediately knew
we had it right, from the big happy smiles on their faces. They, on the spot,
started writing some RGEs on the blackboard and we all got more and more
excited. Very late in the evening John called me at home to assure me that he
had looked at it again and he was convinced that it worked. Let us see how
it works \cite{21}. Consider the RGEs schematically \cite{7}
\begin{equation}
{d\widetilde m^2\over dt}={1\over(4\pi)^2}\left\{ -\sum_{i=1,2,3} c_i g^2_i
M^2_i+\alpha h^2_t(\sum_k\widetilde m^2_k)\right\}
\label{10}
\end{equation}
where $\widetilde m$ generically refers to squarks, sleptons, or Higgs bosons,
$g_{1,2,3}$ are the $SU(3)\times SU(2)\times U(1)$ gauge couplings, $M_i$ are
the corresponding gaugino masses, $h_t$ is the top-quark Yukawa coupling
(ignoring all other Yukawa couplings), and $c_i$ and $\alpha$ are numbers of
order 1, if not zero. It becomes clear now that as $t\equiv\ln Q^2$ decreases
from the Planck scale down to the LEP scale, the different particle masses
evolve differently because the $c_i$ and $\alpha$ depend on their particular
$SU(3)\times SU(2)\times U(1)$ ``charges". Here are the qualitative results
of such an analysis \cite{7}.
\begin{itemize}
\item Sleptons: only $c_{1,2}\not=0$ and thus, as we reach low energies,
$m^2_{\tilde l}$ slightly increases from its $m^2_0>0$ value at Planck-like
scales: $U(1)_{\rm em}$ is safe.
\item Squarks: All $c_{1,2,3},\alpha\not=0$, but still, since $h^2_t\lsim{\cal
O}(g^2_2)$, $m^2_{\tilde q}$ {\em increases} considerably as we reach low
energies, thus always $m^2_{\tilde q}>0$: $SU(3)_{\rm color},U(1)_{\rm em}$ are
safe.
\item Higgs: Here $c_{1,2},\alpha\not=0$ and there is a ``fight" between the
$-g^2_2$ term against the $+h^2_t$ term. Clearly, for big enough $h_t$,
$h^2_t\approx{\cal O}(g^2_2)$, the exciting possibility emerges of {\em
decreasing} $m^2_{\rm Higgs}$, as we reach low energies, enough to hit zero
or even negative values: $m^2_h\le0|_{\rm low\ energy}$, thus causing SB of
the electroweak interactions! Because it is due to radiative corrections, \ie,
of dynamical origin, we called it {\em Radiative Electroweak Symmetry Breaking}
(REWSB).
\end{itemize}

Actually, detailed calculations showed the following highly interesting facts
\cite{7}:
\begin{enumerate}
\item In order to have REWSB, we needed a rather heavy top-quark
$m_t\approx{\cal O}(M_W)$, a rather bold statement for `82, where the top-mass
lower bounds were in the 20 GeV range!
\item We were able to determine {\em dynamically} the electroweak scale by
basically determining the scale $\mu_0$ where $m^2_h$ hits zero:
\begin{equation}
{M_W\over M}\approx {\mu_0\over M}\approx e^{-2\pi/3\bar\alpha_t(1+\bar A^2_t)}
\label{11}
\end{equation}
where $M$ is some scale in the ($M_{\rm GUT}\to M_{Pl}$) region, and
$\bar\alpha_t\equiv\bar h^2_t/4\pi$, $\bar A_t$ are some ``mean" values of
$\alpha_t$ and $A_t$ in the ($M_W\to M$) energy range. Of course, the
similarity with the standard GUT relation
\begin{equation}
{\Lambda_{\rm QCD}\over M}\approx e^{-2\pi\sin^2\theta_W/3\alpha_{\rm em}}
=e^{-2\pi/3\alpha_2}
\label{12}
\end{equation}
is rather striking and certainly not accidental!
\end{enumerate}

A few days after we finished the paper \cite{21}, I presented it at a meeting
in Tokyo. After my talk, a shy Japanese fellow by the name of K. Inoue
approached me and gave me a preprint. I did not look at it until I was airborne
back to Geneva, and I had almost an apoplexy: they (K. Inoue, \etal\ \cite{22})
had done it also, and even in a more detailed and systematic way! Later on,
many people \cite{23} jumped on the band wagon and completed the picture, so
that REWSB became the standard mechanism for SB of the electroweak interactions
for a large part of our community. It will be unfair not to mention here some
seeds of the REWSB idea due to Ib\'a\~nez and Ross \cite{24}. They also relied
on supersymmetry breaking to enforce $m^2_h<0$, but all their game was at the
electroweak scale, \ie, they were not using any RGEs with boundary conditions
at $M$, and only in the framework of {\em global SUSY}.

Clearly REWSB, through (\ref{11}), provides a resolution of the {\em magnitude}
part of the gauge hierarchy problem (\ref{2}). Furthermore, the {\em dynamical}
explanation of SB of electroweak interactions and at the same time the
justification of why $SU(3)_{\rm color}\times U(1)_{\rm em}$ remains unbroken
make REWSB an offer we cannot refuse. It certainly looks like a basic step
towards the construction of a realistic SUGRA model. Nevertheless, there is a
catch! We have {\em tacitly} assumed, as we run the RGEs (\ref{10}), that the
SUSY soft breaking terms ($m_0,m_{1/2},\cdots$) are not far above the Fermi or
electroweak scale. Otherwise we have to ``freeze" the running (\ref{10}) at
some scale $Q\gg M_W$, which may lead to either $SU(2)\times U(1)$ breaking
at the ``wrong" scale or no $SU(2)\times U(1)$ breaking at all! Catastrophic
alternatives indeed! Then, someone may argue that we have tacitly replaced the
{\em magnitude} part of the gauge hierarchy problem (\ref{2}) by the
{\em SUSY breaking scale} problem, \ie, why should
\begin{equation}
{\widetilde m\over M}\approx {\cal O}(10^{-16})\
\label{13}
\end{equation}
It looks like we have to overcome another stumbling block before we are able to
successfully implement the REWSB. No-scale supergravity comes to its rescue.

\section{No-scale supergravity}
\subsection{No-scale Standard model}
The CERN Cafeteria is always a pretty busy place during lunch time, especially
during summer time, when everyone is visiting CERN. A hot noon in the
mid-summer of `83, I had lunch with Costas Kounnas in the CERN Cafeteria, when
we saw Sergio Ferrara looking desperately for a table. We called him over and
started discussing the talks we all had to deliver in a few days at the
European Physical Society High Energy Physics Meeting in Brighton, England.
Sergio was very complimentary on a paper we have just written \cite{25} on the
solution to the Higgs doublet-triplet splitting problem in supergravity, a
rather notorious problem. I vividly remember my reaction to his comments. I
told him that we were very happy with the resolution of this problem, but that
were very {\em unhappy} with the extraneous fine-tuning that is needed in SUGRA
theories so that they meet
the cosmological constant upper bound (\ref{4}). Sergio answered back by saying
that at least we can do it in SUGRA, unlike SUSY theories where SUSY breaking
implies positive cosmological constant! I told him that we needed to do better,
something like a naturally flat potential that does not need fine-tuning, like
the resolution of the Higgs doublet-triplet splitting problem that we all liked
so much. He smiled enigmatically back at me and we left the discussion there.
Very late in the afternoon he came to my office, closed the door and wrote
something on the blackboard. Busy with preparing my transparencies, I paid no
attention to him. Then he said, ``why don't you check the potential in this
case?" It took me a few seconds to check it and indeed it came out {\em
naturally flat}, exactly what I had asked him. I was speechless, it was so
simple and beautiful and we had not thought about this until that moment!
Within the next two days we had the first draft of the paper \cite{1}, while we
diminished our activity on our Brighton talks! Here is what we found \cite{1}.
In N=1 supergravity theories, the effective potential is given by (neglecting
D-terms) \cite{7}
\begin{equation}
V_{\rm SUGRA}=e^G\left[G_i(G^{-1})^i_j G^j-3\right]
\label{14}
\end{equation}
where the K\"ahler function $G=G(T_i,T^*_i)$ is a real function of the fields
$T_i$, and their complex conjugates $T^*_i$, and $G_i=\partial G/\partial T_i$;
$G^j=\partial G/\partial T^*_j$. Notice that the gravitino mass $m_{3/2}$,
whose non-vanishing value at the ground state is a measure of SUGRA breaking,
is given by
\begin{equation}
m_{3/2}=e^{\VEV{G/2}}
\label{15}
\end{equation}
with $\VEV{G}$ denoting the value of $G$ at the minimum of the potential
(\ref{14}). Clearly, one way to have $\Lambda_c=\VEV{V_{\rm SUGRA}}=0$ is
to have $m_{3/2}=0$, \ie, unbroken supergravity, a phenomenologically
unacceptable case. Another way is to arrange things so that
\begin{equation}
\VEV{G_i(G^{-1})^i_j G^j}=3
\label{16}
\end{equation}
which demands an extraneous fine-tuning between different parameters, and thus
highly improbable. Still another way, which is the mathematical expression of
my question to Sergio is to demand
\begin{equation}
G_i(G^{-1})^i_j G^j=3
\label{17}
\end{equation}
as a {\em field identity}! That is to look for a {\em specific} function
$G(T_i,T^*_i)$ that satisfies (\ref{17}) for {\em every value of $T_i$}, thus
getting a {\em naturally vanishing (flat) potential}, without necessarily a
vanishing $m_{3/2}$, \ie, with SB supergravity! What Sergio had written on
my blackboard was
\begin{equation}
G=-3\ln(T+T^*)
\label{18}
\end{equation}
which clearly obeys (\ref{17}), in the simple case of one field, and thus leads
to a flat $V_{\rm SUGRA}\equiv0$ according to (\ref{14}). Actually, in the case
of only one field, the solution (\ref{18}) is the {\em unique} solution of
(\ref{17})! Furthermore, at this classical level the mass of the gravitino
remains undetermined, while non-zero
\begin{equation}
m_{3/2}={1\over\VEV{(T+T^*)^{3/2}}}
\label{19}
\end{equation}
since $\VEV{T+T^*}$ is undetermined! So, here is what we have achieved
\cite{1}: a naturally vanishing cosmological constant, at least at the
classical level, with SB supergravity, {\em but} with the SUGRA breaking scale
($\propto m_{3/2}$) {\em undetermined}. Costas and I found this last part, {\em
undetermined SUGRA breaking at the classical level}, very exciting, because we
somehow foresaw some connection with our REWSB programme discussed in the
previous section. Here things took a strange turn: Eugene Cremmer and Sergio
Ferrara wouldn't even listen about possible connections to the REWSB mechanism,
and we endep up \cite{1} trying to {\em un-flat} the potential so that we got a
fixed value of $m_{3/2}$
at the classical level! Despite all these needless excursions, Costas and I
managed, under their noses, to sneak in some statements at the end of the paper
about what we really believed was happening. Until this day I have not figured
out why the CF part of our team (CFKN) \cite{1} did not want to even listen
about REWSB. Well, they were paving the way for John Ellis. John has never been
accussed of not getting instantly an important physical point and true to his
reputation, when he saw our (CFKN) paper \cite{1}, he also figured out that we
had in our hands a
mechanism for fixing dynamically all mass scales. He called from Munich, where
he was at a summer school, and eagerly asked me not to write anything further
before he was back at CERN! During this phone conversation the name {\em
no-scale supergravity} was invented, as well as the specific generic scenario
for dynamical determination of all mass scales. Again the idea is very simple.
Applied supergravity usually provides us with some {\em fixed} arbitrary
boundary conditions for the soft breaking parameters at, or close to, the
Planck scale, assuming that all fields are at their minima, as shown in
(\ref{9}). In no-scale supergravity though, as depicted by (\ref{19}), the soft
breaking parameters are {\em not fixed}, at least not at the classical level,
at, or close to, the Planck scale, thus replacing (\ref{9}) by
\begin{equation}
\widetilde m=\widetilde m(T_i)
\label{20}
\end{equation}
and thus we have to rely, once more, on radiative corrections to fix
$\VEV{T_i}$, and thus providing a {\em dynamical determination} of
($m_0,m_{1/2},\cdots$)! Very similar to the way we determined the v.e.v. of
the Higgs field, or equivalently $M_W$ (see (\ref{11})), in the standard model
by the REWSB mechanism. But here we are facing a problem. The $T_i$ fields
are usually singlets under $SU(3)\times SU(2)\times U(1)$, so how can we use
strong and electroweak radiative corrections to fix their v.e.v.'s, a la the
Higgs v.e.v.'s? Well, there is an extra small but profound step we have to take
beyond the REWSB mechanism in order to complete our {\em dynamical
determination} of the SUSY breaking scale. This extra step was taken by John
Ellis, Thanasis Lahanas, Kyriakos Tamvakis, and myself about a month after the
original CFKN paper \cite{1}, by constructing the first ever genuine realistic
no-scale supergravity standard model \cite{2}. Of course, in this realistic
case the no-scale K\"ahler function (\ref{18}) will have additional pieces
representing the ``observable" fields (quarks, Higgs, etc.), but they don't
change at all our basic picture. The electroweak potential at the ``observable"
fields minimum, after the REWSB has occured, takes the generic form \cite{2,7}
\begin{equation}
(V_{\rm E-W})_{\rm min}=-C m^4_{3/2}(T)\  \ln^2\left({\kappa_0
m^2_{3/2}(T)\over\mu^2_0}\right)
\label{21}
\end{equation}
where within $C$ we have absorbed all quantities that do not depend on
$m_{3/2}$, $\kappa_0$ is a constant of order 1, and $\mu_0$ is given by
(\ref{11}). Clearly, $m_{3/2}=m_{3/2}(T)$, as given by (\ref{19}), is an
undetermined {\em dynamical variable}. Thus, we should minimize $(V_{\em
E-W})_{\rm min}$ {\rm further} with respect to the field $T$ or equivalently
with respect to $m_{3/2}(T)$, \ie, we are looking for the {\em minimum
minimorum} of the electroweak potential
\begin{equation}
{d (V_{\rm E-W})_{\rm min}\over dm_{3/2}}=0
\label{22}
\end{equation}
which I have dubbed the {\em no-scale condition}. Applying this general
no-scale condition (\ref{22}) in the specfic case of $(V_{\rm E-W})_{\rm min}$,
as given by (\ref{21}), one determines {\em dynamically} $m_{3/2}$
\begin{equation}
{m^2_{3/2}\over \mu^2_0}={1\over\kappa_0 e}={\cal O}(1)
\label{23}
\end{equation}
and thus by (\ref{19}), fixing {\em dynamically} $\VEV{T+T^*}$.

This is the no-scale answer to the {\em SUSY breaking scale} problem
(\ref{13}). Thus, putting together (\ref{11}) and (\ref{23}) we get
\begin{equation}
{M_W\over M}\approx{m_{3/2}\over M}\approx e^{{\cal O}(-1/\alpha_t)}
\label{24}
\end{equation}
\ie, a {\em dynamical determination} of the fundamental scales of high-energy
physics, the electroweak scale and the SUSY breaking scale. We also understand
intuitively why they are close together: after all, in the no-scale framework,
it is the physics of the electroweak potential that finally fixes $m_{3/2}$,
thus one a priori expects $m_{3/2}\approx{\cal O}(M_W)$, what else is there?

Actually, there is more to our story. As originally observed in CFKN \cite{1},
and elaborated in great detail in EKN(I) \cite{3}, the flatness of the
potential, when one
used the no-scale K\"ahler form (\ref{18}) is not just a caprice of the
relevant equations, but it is due to some deep symmetry reasons. The flatness
of the potential has its origin in the existence of a non-compact symmetry,
$SU(1,1)$, which contains as subgroups, imaginary translations ($T\to
T+i\beta$), dilatations,
and conformal transformations. We then discovered, to our delight, that the
$SU(1,1)$ invariant ``hidden" sector of our theory was {\em isomorphic} to
the scalar sector of the N=4 extended supergravity. We could trace back the
well-known properties of the absence of a scalar potential in the N=4 theory,
to its {\em possession} of the non-compact $SU(1,1)$ invariance. The absence
of a scalar potential and the vanishing of the cosmological constant in higher
$N>4$ (ungauged) supergravities are linked to their larger non-compact global
invariance groups which contain $SU(1,1)$ as a subgroup. In other words, our
choice of the no-scale K\"ahler function (\ref{18}) was not hard to justify
as it had its {\em dynamical origin} in $N\ge4$ supergravity which, at least
at that time, were popular \cite{26} for providing a fundamental theory of the
Planck scale! An extra bonus, discussed in detail in EKN(I) \cite{3}, was the
fact that, due to the $SU(1,1)$ non-compact symmetry, the {\em observable}
low-mass fields have to be combined with the {\em hidden sector} in a
non-trivial way if there is to be a non-trivial limit as $m_{3/2}/M_{Pl}\to0$.
Light fields {\em should have non-zero} conformal weights if their
superpotential terms are not to vanish when $m_{3/2}/M_{Pl}\to0$. The Yukawa
couplings ($h_i$) rescale differently, depending on their corresponding
``conformal ($\subset SU(1,1)$) weights"
\begin{equation}
{m_{f_i}\over M_W}\approx{\cal O}\left({M_W\over M}\right)^{\lambda_i},
\qquad i=1,2,3\ ({\rm generation\ index})
\label{25}
\end{equation}
thus providing the ``seeds" for a resolution of the fermion mass hierarchy
problem (\ref{3}). For example, pick up $\lambda_1\approx1/8$,
$\lambda_2\approx1/4$, $\lambda_3\approx0$, rather normal values for the
``weights", and we have the gross features of (\ref{3}) explained. Of course,
it remains to the fundamental theory, that would contain our no-scale
structure, to fix these specific ``weights" in a satisfactory way.

It is remarkable that the dynamical structure of the no-scale supergravity
is such that not only solves satisfactorily both facets of the gauge hierarchy
problem, as seen in (\ref{24}), but it provides the ``seeds" for resolving the
fermion mass hierarchy problem, as seen in (\ref{25}). The common root of the
resolution of all the above problems lies closely in the flatness of the
potential, \ie, a naturally vanishing cosmological constant, at least at the
classical level, due to the existence of a {\em dynamically derived} (in
$N\ge4$ extended SUGRAs) non-compact, global $SU(1,1)$ symmetry. Thus, the
deep correlation (\ref{5}), which is the focal point of no-scale supergravity.

It should be stressed that in order for the no-scale programme to work, we
have to {\em assume} \cite{3,7} that ${\rm Str}\,{\cal
M}^2=\sum_j(-1)^{2j}(2j+1){\rm Tr}\,{\cal M}^2_j$ {\em vanishes}. Otherwise
there will be an extra term
$({\rm Str}\,{\cal M}^2)\Lambda^2\approx m^2_{3/2}M^2_{Pl}$ in $V_{\rm eff}$
that would imply either $m_{3/2}=0$ or $M_{Pl}$! In other words, there would
be {\em no} no-scale framework if ${\rm Str}\,{\cal M}^2\not=0$. From the
first days of no-scale models we stressed this point repeatedly (see \eg,
a huge footnote on p. 408 in EKN(I) \cite{3}), but without having an {\em
explicit} fundamental theory at the Planck scale, we simply included the ${\rm
Str}\,{\cal M}^2=0$ condition in the {\em desiderata} of the correct unified
model. It turns out that this specific {\em desideratum} is a tough
constraint to meet, at least in presently popular string models.

\subsection{No-scale GUTs}
Costas Kounnas is usualy in an ``excited state". In the winter of `82-83 he was
almost ``ionized". Toward the end of our work on the first EKN paper \cite{3},
he succeeded to generalize the no-scale K\"ahler potential (\ref{18}) in a
non-trivial way, for $N>1$ chiral fields, by utilizing $SU(N,1)$ non-compact
global symmetry instead of $SU(1,1)$. He also observed that in this case, one
gets (\ref{8}) but without the ``soft-breaking" terms, \ie, $m_0=A=B=0$!
For John and me that was the wrong way to go. Who needed unbroken global SUSY
theories? Costas, charmed by the nice geometrical features of the $SU(N,1)$
theory, wouldn't hear any criticism. He was talking continously about $SU(N,1)$
to the extent that at any time I would see him in the corridor, I would hide
in someone's office. During the Christmas/New Year holidays of `83-84, while
I was in Athens, Greece, I started thinking about no-scale GUTs. Unfortunately,
almost immediately I stopped on my ``tracks" by very unpleasant thoughts. We
had a rather severe problem to solve. In the general case when there is an
intermediate scale in the theory, like $M_G$, the GUT scale usually taken to
be ${\cal O}(10^{16}\GeV)\,(\ll M_{Pl})$, there is a contribution to the
effective potential of the form
\begin{equation}
\Delta V\approx\left((M^2_G+m^2_{3/2})^2-M^4_G\right)\ln\left({m^2_{3/2}\over
M^2_G}\right)\approx {\cal O}(m^2_{3/2}M^2_G)\ln\left({m^2_{3/2}\over
M^2_G}\right)\ ,
\label{26}
\end{equation}
which, by following the steps that led to (\ref{23}), implies
$m_{3/2}\approx{\cal O}(M_G)$, not a very desirable result! Clearly, we have to
make sure that terms ${\cal O}(m^2_{3/2}M^2_G)$ do not appear in $\Delta V$,
but how? Well, the only {\em natural} way that we know how to do this is to
employ {\em unbroken} global SUSY for the {\em heavy sector}. And then it
clicked. Suddenly, it occured to me that what Costas had found ($m_0=A=B=0$)
was the solution to our severe problem. Indeed, after my return to CERN, we
worked out the basic framework that enabled us to construct no-scale GUTs
\cite{4}. Here are the main points. The no-scale K\"ahler potential for
N-chiral fields (\eg, the hidden sector $T$-field and N-1 ``observable" fields
$\psi_i$) is given by \cite{4} (neglecting the superpotential term for
simplicity)
\begin{equation}
G=-3\ln(T+T^*-\sum_i\psi_i\psi_i^*)
\label{27}
\end{equation}
which has the following remarkable properties:
\begin{itemize}
\item Naturally vanishing cosmological constant, \`a la (\ref{17}), due to the
existence of an $SU(N,1)\supset SU(1,1)$ non-compact global symmetry available
in $N\ge5$ extended supergravities \cite{26}.
\item $m_0=A=B=0$, directly related to the existence of the $SU(N,1)$
non-compact symmetry, thus averting the disastrous ${\cal O}(m^2_{3/2}M^2_G)$
{\em heavy sector} contribution to the effective potential.
\end{itemize}

On the other hand, we had at our disposal the possibility of turning on
low-energy sector gaugino masses ($m_{1/2}\not=0$), thus providing global SUSY
breaking in the low-energy sector, while keeping the {\em heavy sector}
supersymmetric. One has to be careful that radiative corrections do not filter
into the {\em heavy sector} any SUSY breaking, and this is guaranteed if we
push $M_G\approx{\cal O}(M_{Pl})$, and thus avoiding any renormalization of the
heavy sector. We decided to go this way \cite{4}, and we relied on the future
TOE to explain dynamically our assumptions:
\begin{itemize}
\item $m_{1/2}\not=0$ and $M_G\approx{\cal O}(M_{Pl})$.
\end{itemize}

Actually, we were bold enough to enlarge the spectrum of the SUSY $SU(5)$ by an
extra $\bf 10+\overline{10}$, so that the usual case of $M_G\approx10^{16}\GeV$
is replaced by $M_G\approx{\cal O}(M_{Pl})$! A rather daring and provocative
proposal in 1984! Clearly, the idea and its realization, of a GUT theory with
$M_G\approx{\cal O}(M_{Pl})$ is not {\em new} and it has {\em not} been
conceived only {\em a posteriori} in order to reconcile the LEP data indicating
$M_G\approx{\cal O}(10^{16}\GeV)$, under the assumption of minimal SUSY $SU(5)$
spectrum, with the string scale $M_S\approx{\cal O}(10^{18}\GeV)$.

Before ending this section, just a few words about some important subsequent
developments of no-scale GUTs that unfortunately I have no time or space to
cover.
\begin{itemize}
\item We found that the condition $m_{3/2}\approx{\cal O}(M_W)$ could be
abandoned and $m_{3/2}$ could be anything, \eg, $m_{3/2}\approx{\cal
O}(M_{Pl})$ \cite{5} or $m_{3/2}\le{\cal O}(1\,{\rm KeV})$ \cite{6}, as long as
$m_0,m_{1/2}\approx{\cal O}(M_W)$. In certain cases there is a decoupling of
$m_{3/2}$, and in everything I discussed above, its role is played by
$m_{1/2}$. This is what we dubbed with Thanasis Lahanas, the ``Gravitino
Liberation Movement" (GLM) in our Physics Report review of no-scale
supergravity \cite{7}, conceived in a liquid lunch in Ferney-Voltaire, a
village in France, near CERN.
\item With John Ellis and Kari Enqvist we entertained the very interesting
possibility that a similar mechanism like the one that fixes
$m_{3/2}=m_{3/2}({\rm Re}\,T)$, may be at work to fix dynamically $\theta_{\rm
QCD}=0$. Indeed, we worked out examples \cite{6} where $\theta_{\rm
QCD}=\theta_{\rm QCD}({\rm Im}\,T)$, while QCD nonperturbative dynamics provide
a potential to the otherwise decoupled ${\rm Im}\, T$ field, and thus ensuring
that at its minimum  $\theta_{\rm QCD}=0$, according to the well-known
Peccei-Quinn theorem \cite{27}.
\end{itemize}

While clearly no-scale supergravity has some very appealing and unique
properties, the big question remains: Is there any fundamental theory, at the
Planck scale, that has no-scale supergravity as its infrared limit? The answer
is emphatically yes, and the fundamental theory is {\em string theory} (ST).

\section{String no-scale supergravity}
In February of `85, and while a bunch of theorists including John Ellis and
myself had lunch at the CERN cafeteria, we got some very exciting news.
Gabriele Veneziano joined us, and as he was, literally, sitting down he broke
the news ``I just got a paper from Ed (Witten) that contains a derivation of
no-scale
supergravity from superstring theory". I still remember the Cheshire cat's
smile  on John's face, and my infinite impatience to finish lunch and get
Ed's paper \cite{28}. Using an ``educated" dimensional reduction, highly
imitative of compactification on a, then popular, Calabi-Yau manifold, Witten
succeeded in deriving the following form for the K\"ahler function \cite{28}
(again neglecting, for simplicity, the superpotential term)
\begin{equation}
G=-3\ln(T+T^*-\sum_i\psi_i\psi^*_i)-\ln(S+S^*)
\label{28}
\end{equation}
which of course, is nothing else, but our no-scale GUTs K\"ahler function
(\ref{27}), amended by the inclusion of an extra {\em no-scale-type term},
corresponding to the dilaton ($S$) field endemic in string theory.
In this case the K\"ahler manifold is ${SU(N,1)\over SU(N)\times U(1)}\otimes
{SU(1,1)\over U(1)}$, strongly characteristic of the no-scale framework. Not
even in my wildest dreams I would have thought that we would have a
``derivation" of our no-scale framework within a year of its discovery, from
a supposedly and purportedly fundamental theory, like string theory.

Actually, the derivation of an effective supergravity theory from string theory
has become a special branch of research in string theory. It should be stressed
that Witten's initial studies \cite{28} have been replaced by much more
sophisticated techniques, that have led to {\em specific, exact} forms of the
K\"ahler function, always involving non-compact symmetries and always leading
to some type or other of no-scale structure. The first complete studies in
$D=4$ superstring theories were done \cite{29} in the framework of the
fermionic formulation (FFF) \cite{30} and led to very interesting
phenomenology. For the most recent study see Ref.~\cite{31}, where all previous
work is fairly completely mentioned.

The basic question that usually arises is, what is the physical reason that
string theory provides a no-scale framework as its infrared limit? There is a
simple reason for all this as follows:
\begin{description}
\item (i) At the level of the space-time effective action, we shouldn't forget
that we start with something like $N=1$ SUGRA in $D=10$, which can be seen as
$N=4$ SUGRA in $D=4$, and thus our space-time effective action should, at some
level, contain among its symmetries the non-compact symmetries characteristic
of extended supergravities, and as discussed in the previous section, the
raison d'etre of the no-scale structure. Note that in the case of (FFF)
\cite{30}, we always start with $N=4$ SUGRA in $D=4$, thus the above comments
cover also this case.
\item (ii) At the level of the 2-d world-sheet action, we shouldn't forget that
conformal invariance entails the vanishing of the 2-d world-sheet
$\beta$-functions $\beta_\phi$, which in turn provide the Equations of Motion
(EOM) for the $D=4$ physical fields $\phi$
\begin{equation}
\beta_\phi={d V_{\rm eff}\over d\phi}=0\ .
\label{29}
\end{equation}
For certain fields, called moduli (like the $T$ and $S$ discussed above), the
vanishing of the $\beta$-functions (or $dV_{\rm eff}/d\phi$) does not occur
only at a critical point, but at a critical line, thus providing {\em flat
directions} for these fields, at least in perturbation theory
\begin{equation}
\left.{dV_{\rm eff}\over d\phi}\right|_{\phi\in{\rm critical\ line}}=0\ .
\label{30}
\end{equation}
But, {\em flat directions} are the logo of the no-scale framework, and we found
here that they have their roots {\em deep} into the basics of string theory!
\end{description}

Of course, one expects much
more sophisticated no-scale K\"ahler structures \cite{31}, that we envisioned
initially \cite{3,4}, since here there is a multitude of moduli fields and
after all, in string theory the no-scale structure is {\em dynamically
derived}, it is not intuitively imposed. You get, what you get! Actually, this
is good news because
now we can address dynamically some of the assumptions that we have made
before, including the vanishing of ${\rm Str}\,{\cal M}^2$, and the need for
$m_0=A=B=0$; $m_{1/2}\not=0$ and $M_G\approx{\cal O}(M_{Pl})$ in the case of
no-scale GUTs. Indeed, there is a very simple, ``realistic" example that shows
how one can avert the ${\rm Str}\,{\cal M}^2\not=0$ problem. Consider the case
of 3-singlets (moduli) fields $T,S,U$ and $N-1$ ``charged" fields $\psi_i$,
entering the K\"ahler function $G$ in the following way \cite{32} (including
the superpotential term $W$)
\begin{equation}
G=-3\ln(T+T^*-\sum_i\psi_i\psi^*_i)-\ln(S+S^*)-\ln(U+U^*)+\ln|W(\psi,S,U)|^2
\label{31}
\end{equation}
corresponding to a stringy-inspired K\"ahler manifold ${SU(N,1)\over
SU(N)\times U(1)}\times {SU(1,1)\over U(1)}\times{SU(1,1)\over U(1)}$. Clearly
this looks very familiar (see (\ref{27}),({\ref{28})) and the extra field $U$
makes nothing to change the good features mentioned above. On the other hand,
it is tremendously helpful in cancelling the ${\rm Str}\,{\cal M}^2$ term.
Indeed, using the techniques developed in EKN(II) \cite{4} one easily finds
\cite{32}
\begin{equation}
{\rm Str}\,{\cal M}^2\propto (N+2)-1-\left({(N+1)\over3}G_TG^T\right)=0
\label{32}
\end{equation}
since $G_TG^T=3$ from the vacuum energy cancellation. Thus, our basic
assumption of ${\rm Str}\,{\cal M}^2=0$ can be met in realistic
string-inspired/derived no-scale models. It shouldn't escape our notice the
deep connection between the vanishing of ${\rm Str}\,{\cal M}^2$ and the
vanishing of the cosmological constant. Concerning the highly desirable
boundary conditions $m_0=A=B=0$; $m_{1/2}\not=0$, they are endemic in string
theories, and all kinds of different approximations usually yield the above
no-scale boundary conditions.
Concerning the $M_G\approx{\cal O}(M_{Pl})$, it is enough to remind ourselves
that string theory provides ``free" gauge coupling unification at a {\em
dynamically determined} scale $M_{\rm string}\approx{\cal
O}(6\times10^{17}\GeV)$ thus making the identification $M_G\approx{\cal
O}(M_{\rm string})$ rather natural. It is left upon to some extra
representations, amending the minimal SUSY Standard Model, to push the {\em
apparent} LEP unification scale of ${\cal O}(10^{16}\GeV)$ close to the string
scale, $M_{\rm string}\approx{\cal O}(6\times10^{17}\GeV)$, exactly as
envisaged in EKN(II) \cite{4} more than ten years ago.

It is very encouraging that our basic assumptions in building the no-scale
framework may find a natural explanation in string theory. Furthermore, it is
easy to see that in string no-scale supergravity we have the potential for
calculating basically from first principles dynamically all relevant
parameters. Here are the basic steps:
\begin{itemize}
\item $g^2=1/\VEV{{\rm Re}\, S}$, the string unification gauge coupling is
given in terms of $\VEV{{\rm Re}\,S}$, introduced in (\ref{28}), that is
expected to
be fixed dynamically, presumably non-perturbatively, but still one naturally
expects $\VEV{{\rm Re}\,S}/M_{\rm string}\approx{\cal O}(1)$, thus
$g^2\approx{\cal O}(1)$ as indicated phenomenologically. Notice that once we
have determined dynamically the value of $g^2\,(\approx{\cal O}(1))$ and the
string scale $M_{\rm string}\approx{\cal O}(6\times10^{17}\GeV)$, we can
determine {\em dynamically} the values of $\alpha_3,\alpha_2,$ and $\alpha_{\rm
em}$ at the LEP scale!
\item $h_i|_{Q\approx M_{\rm string}}\approx g(\VEV{\phi}/M_{\rm string})^n$,
the Yukawa couplings are given in terms of $g$, and some combination of VEVs
of appropriate singlets \cite{33}. Clearly, $\VEV{\phi}/M_{\rm
string}\approx{\cal O}(1/10)$, as indicated in many specific string models, and
$n=0,1,2$, corresponding to 3rd, 2nd, and 1st generation respectively,
``reproduces" with sufficient accuracy the observed fermion mass spectrum
\cite{34}. Note that $h_t\approx {\cal O}(g)$, as observed \cite{35} and as
required by REWSB mechanism. As an example
of this programme, let me mention a rather amusing, interesting, and novel
relation that we got with Jorge Lopez back in 1990 \cite{36}
\begin{equation}
{m_c\over m_t}\sim{1\over2}\left({m_e\over m_\tau}\right)^{1/2}
\label{33}
\end{equation}
which implies a top-quark mass ${\cal O}(160-170\GeV)$! Incidentally, detailed
studies \cite{37}, by Jorge Lopez, Nino Zichichi, and myself, of the top-quark
Yukawa coupling in string theory ($h_t\approx {\cal O}(g)$), enabled us to make
a firm prediction of $m_t\in[150-180\GeV]$ range, just a few months before the
offical FNAL/CDF announcement \cite{35}.
\item $\widetilde m/M_{\rm string}\approx {m_W\over M_{\rm string}}\sim{\cal
O}(e^{-1/\alpha_t})$, by employing the no-scale mechanism with a ``heavy" top
quark dynamically derived in the previous step.
\end{itemize}

Thus, {\em string no-scale supergravity} may provide us with a framework where
basically all physically relevant quantities/parameters would be dynamically
determined, \ie, a Theory of Everything (TOE) with the ability of
{\em Dynamical Determination Of Everything} (DDOE). Nevertheless, we shouldn't
get carried away by the phenomenal success of the good ``mix" of string theory
and the no-scale mechanism. Even, if, which I personally doubt, it turns out
that, string theory was an ephimeral illusion, the no-scale supergravity will
be always there, available as an effective theory, able to give a good
description of reality at energies below the Planck scale.

Until now I have deliberately avoided any discussion on the value of the
cosmological constant ($\Lambda_c$) beyond the classical level. Here, I will
try to explain my attitude and try to answer some skepticism that might have
been born concerning the physics of the no-scale models. We started by
demanding naturally a zero tree-level cosmological constant and this was
related to the $SU(1,1)$ non-compact symmetry, but then with the introduction
of the observable sector the vacuum energy becomes non-zero $V_{\rm E-W}=-{\cal
O}(M^4_W)\not=0$ (\ref{21}). This results in a cosmological constant
$\Lambda_c=10^{-60}M^4_{Pl}$, which is much greater than the one actually
observed (\ref{4}) but still very small in comparison with an energy
whose natural order is expected to be ${\cal O}(m^2_{3/2}M^2_{Pl})$. At this
point, one should not forget that we were dealing with an effective low-energy
theory. In other words, we considered the tree-level gravitational interactions
and then took the flat limit $M_{Pl}\to\infty$. In this way, at low energy
one completely forgets the gravitational interactions and considers only the
electroweak and strong forces. One cannot expect to find vanishing
contributions to the cosmological constant from only the radiative corrections
of the non-gravitational interactions. The vanishing of the cosmological
constant is a matter that concerns the gravitational interactions and these
were ignored at scales $\mu\ll M_{Pl}$ by letting $M_{Pl}\to\infty$ (flat
limit). The full theory should give $\Lambda_c<10^{-120}M^4_{Pl}$ when
gravitational corrections are taken into account, but at present we do not
know how to do that. Hopefully, string no-scale supergravity will succesfully
resolve this problem.

During the last few years, with my close collaborators Jorge Lopez and Nino
Zichichi, both firm believers and strong practitioners of no-scale
supergravity, we have developed a framework of studying string no-scale
supergravity from its lofty string origin down to the nitty-gritty details of
its experimental consequences \cite{38}. Only then, we believe, we can have the
complete theory and we will be able to ``prove" or refute it. We believe that
we are not far from developing a genuine, experimentally accessible, {\em
no-parameter stringy no-scale model} that will be an explicit realization of
the basic steps I sketched above. If we are lucky enough, maybe someone of us
will give you the details of how it happened, at the next History conference of
this series.

\section*{Acknowledgments}
It is a great pleasure for me to express my sincere thanks to the
co-discoverers of the no-scale supergravity (in alphabetical order): E.
Cremmer, J. Ellis, S. Ferrara, C. Kounnas, A. Lahanas, and K. Tamvakis for the
great, exciting, and somehow singularly unique times that we shared together.
Many thanks also are due to my close collaborators Jorge Lopez and Nino
Zichichi for not only keeping the focus and the spirit of no-scale supergravity
very well alive, but also for bringing it to new heights, by digging deeper and
deeper into its structure. Finally, I would like to express my deep gratitude
to Harvey Newman and Tom Ypsilantis, for an impecably organized conference and
for giving me the opportunity to address such an exceptionally talented
audience, including lots of heroes of my youth. This work has been supported in
part by DOE grant DE-FG05-91-ER-40633.

\end{document}